\documentclass[11pt]{4ssmmp} %%% Please do not change this class !!! %%%

\usepackage{fancyhdr}    % headings style
\usepackage{graphicx}
\usepackage{rotating}
\renewcommand{\headrulewidth}{0.4pt}

\begin{document}
 \baselineskip=11pt

\title{Initial Singularity, $\Lambda$-Problem and Crossing the Phantom Divide
in Scale Invariant TMT Model\hspace{.25mm}}
\author{\bf{E. I. Guendelman}\hspace{.25mm}\thanks{\,e-mail address:
guendel@bgu.ac.il  }  \normalsize{} \vspace{2mm}  \bf{and A.  B.
Kaganovich}\hspace{.25mm}\thanks{\,e-mail address:
alexk@bgu.ac.il}
\\ \normalsize{Physics Department, Ben Gurion University, Beer
Sheva 84105, Israel}  }

\date{}

\maketitle

\begin{abstract}
\vspace{2cm} % delate this line

In the framework of the scale invariant model of the Two Measures
Field Theory (TMT), we study the dilaton-gravity sector in the
context of spatially flat FRW cosmology. The scale invariance is
spontaneously broken due to the intrinsic features of the TMT
dynamics.  If no fine tuning is made, the effective
$\phi$-Lagrangian $p(\phi,X)$ depends quadratically upon the
kinetic term $X$. Hence TMT represents an explicit example of {\it
the effective} $k$-{\it essence resulting from first principles
without any exotic term} in the underlying action intended for
obtaining this result. Depending of the choice of regions in the
parameter space (but without fine tuning), TMT exhibits
interesting outputs for cosmological dynamics, for example: a)
{\it Absence of initial singularity of the curvature while its
time derivative is singular}. This is a sort of "sudden"
singularities studied by Barrow on purely kinematic grounds. b)
Power law inflation in the subsequent stage of evolution which
ends with a {\it graceful exit}  into the state with zero
cosmological constant (CC). c) Possibility of {\it resolution of
the old CC problem}. From the point of view of TMT, it becomes
clear why the old CC problem cannot be solved (without fine
tuning) in conventional field theories;  d) There is a wide range
of the parameters such that in the late time universe: the
equation-of-state  $w=p/\rho <-1$; $w$ {\it asymptotically} (as
$t\rightarrow\infty$) {\it approaches} $-1$ {\it from below};
$\rho$ approaches a constant, the smallness of which does not
require fine tuning of dimensionfull parameters.
\end{abstract}



\newpage

 {\section{Basis of Two Measures Field Theory}

TMT is a generally coordinate invariant theory where {\it all the
difference from the standard field theory in curved space-time
consists only of the following three additional assumptions}:

1. The main supposition  is that for describing the effective
action for 'gravity $+$ matter' at energies below the Planck
scale, the usual form of the action $S = \int L\sqrt{-g}d^{4}x$ is
not complete. Our positive hypothesis is that the effective action
has to be of the form\cite{GK1}
\begin{equation}
    S = \int L_{1}\Phi d^{4}x +\int L_{2}\sqrt{-g}d^{4}x
\label{S}
\end{equation}
 including two Lagrangian densities $ L_{1}$ and $L_{2}$ and two
measures of integration $\sqrt{-g}$ and $\Phi$. One is the usual
measure of integration $\sqrt{-g}$ in the 4-dimensional space-time
manifold equipped with the metric
 $g_{\mu\nu}$.
Another  is the new measure of integration $\Phi$ in the same
4-dimensional space-time manifold. The measure  $\Phi$ being  a
scalar density and a total derivative may be defined
 by means of  four scalar fields $\varphi_{a}$
($a=1,2,3,4$):
\begin{equation}
\Phi
=\varepsilon^{\mu\nu\alpha\beta}\varepsilon_{abcd}\partial_{\mu}\varphi_{a}
\partial_{\nu}\varphi_{b}\partial_{\alpha}\varphi_{c}
\partial_{\beta}\varphi_{d}.
\label{Phi}
\end{equation}}

2. It is assumed that the Lagrangian densities $ L_{1}$ and
$L_{2}$ are functions of all matter fields, the dilaton field, the
metric, the connection
 but not of the
"measure fields" $\varphi_{a}$. In such a case, i.e. when the
measure fields  enter in the theory only via the measure $\Phi$,
  the action (\ref{S}) possesses
an infinite dimensional symmetry\cite{GK1}). One can hope that
this symmetry should prevent emergence of a measure fields
dependence in $ L_{1}$ and $L_{2}$ after quantum effects are taken
into account.

3. Important feature of TMT that is responsible for many
interesting and desirable results of the field theory models
studied so far\cite{GK1}
 consists of the assumption that all fields, including
also metric, connection  and the {\it measure fields}
$\varphi_{a}$ are independent dynamical variables. All the
relations between them are results of equations of motion.  In
particular, the independence of the metric and the connection
means that we proceed in the first order formalism and the
relation between connection and metric is not necessarily
according to Riemannian geometry.

We want to stress  that except for the listed three assumptions we
do not make any changes as compared with principles of the
standard field theory in curved space-time. In other words, all
the freedom in constructing different models in the framework of
TMT consists of the choice of the concrete matter content and the
Lagrangians $ L_{1}$ and $L_{2}$ that is quite similar to the
standard field theory.

Variation of the measure fields $\varphi_{a}$ results in  the
equation of the form $B^{\mu}_{a}\partial_{\mu}L_{1}=0$ where the
matrix $B^{\mu}_{a}$ is non-degenerate  if $\Phi\neq 0$. Then we
get
\begin{equation}
 L_{1}=sM^{4} =const
\label{varphi}
\end{equation}
where $s=\pm 1$ and $M$ is a constant of integration with the
dimension of mass. In what follows we make the choice $s=1$.

Applying the Palatini formalism in TMT one can show  that in
addition to the Christoffel's connection coefficients, the
resulting relation between connection and metric includes also the
gradient of the ratio of the two measures
\begin{equation}
\zeta \equiv\frac{\Phi}{\sqrt{-g}} \label{zeta}
\end{equation}
which is a scalar field. Consequently geometry of the space-time
with the metric $g_{\mu\nu}$ is non-Riemannian. The gravity and
matter field equations obtained by means of the first order
formalism contain both $\zeta$ and its gradient. It turns out that
at least at the classical level, the measure fields $\varphi_{a}$
affect the theory only through the scalar field $\zeta$. The
consistency condition of equations of motion has the form of a
constraint which determines $\zeta (x)$ as a function of matter
fields. By an appropriate change of the dynamical variables which
includes a  transformation of the metric, one can formulate the
theory in a Riemannian  space-time. The corresponding frame we
call "the Einstein frame". The big advantage of TMT is that in the
very wide class of models, {\it the gravity and all matter fields
equations of motion take canonical GR form in the Einstein frame}.
 All the novelty of TMT in the Einstein frame as compared
with the standard GR is revealed only
 in an unusual structure of the scalar fields
effective potential (produced in the Einstein frame), masses of
fermions  and their interactions with scalar fields as well as in
the unusual structure of fermion contributions to the
energy-momentum tensor: all these quantities appear to be $\zeta$
dependent\cite{GK1}-\cite{GK3}. This is why the scalar field
$\zeta (x)$ determined by the constraint as a function of matter
fields, has a key role in the dynamics of TMT models.

\section{Scale invariant model}

Our TMT model includes gravity and a dilaton field $\phi$. In the
early universe $\phi$ plays the role of inflaton and in the late
universe it is is a part of the dark energy.
 We postulate  the global scale
invariance:
\begin{equation}
    g_{\mu\nu}\rightarrow e^{\theta }g_{\mu\nu}, \quad
\Gamma^{\mu}_{\alpha\beta}\rightarrow \Gamma^{\mu}_{\alpha\beta},
\quad \varphi_{a}\rightarrow \lambda_{ab}\varphi_{b}, \quad
\phi\rightarrow \phi-\frac{M_{p}}{\alpha}\theta , \label{st}
\end{equation}
where $\det(\lambda_{ab})=e^{2\theta}$ and the Lagrangian $L_1$,
$L_2$ in the action (\ref{S}) are:
\begin{eqnarray}
L_1=e^{\alpha\phi /M_{p}}\left[-\frac{1}{\kappa}R(\Gamma
,g)+\frac{1}{2}g^{\mu\nu}\phi_{,\mu}\phi_{,\nu}-V_{1}e^{\alpha\phi
/M_{p}}\right]\nonumber\\
L_2=e^{\alpha\phi /M_{p}}\left[-\frac{b_g}{\kappa}R(\Gamma
,g)+\frac{b_{\phi}}{2}g^{\mu\nu}\phi_{,\mu}\phi_{,\nu}-V_{2}e^{\alpha\phi
/M_{p}}\right]
 \label{totaction}
\end{eqnarray}
One should point out that there are two types of the gravitational
terms and
 of the "kinetic-like terms"  which
respect the scale invariance : the terms of the one type coupled
to the
 measure $\Phi$ and those of the other type
coupled to the measure $\sqrt{-g}$. Using the freedom in
normalization of the measure fields $\varphi_{a}$   we set the
coupling constant of the scalar curvature to the measure $\Phi$ to
be  $-\frac{1}{\kappa}$. Normalizing all the fields such that
their couplings to the measure $\Phi$ have no additional factors,
we are not able in general to provide the same in the terms
describing the appropriate couplings to the measure $\sqrt{-g}$.
This fact explains the need to introduce the dimensionless real
parameters $b_g$ and $b_{\phi}$ and we will only assume that {\it
they are positive and have close orders of magnitudes}. Note that
in the case of the choice $b_g=b_{\phi}$ we would proceed with
{\it the fine tuned model}. The real positive parameter $\alpha$
is assumed to be of the order of unity; in all numerical solutions
presented in this talk we set $\alpha =0.2$. For the Newton
constant we use $\kappa =16\pi G $, \, $M_p=(8\pi G)^{-1/2}$. Note
also the possibility of introducing two different pre-potentials
which are exponential functions of the dilaton $\phi$ coupled to
the measures $\Phi$ and $\sqrt{-g}$ with factors $V_{1}$ and
$V_{2}$  respectively. Such $\phi$-dependence provides the scale
symmetry (\ref{st}). However $V_{1}$ and $V_{2}$ might be
Higgs-dependent and then they play the role of the Higgs
pre-potentials\cite{GK2-Higgs}.

 The common feature of all equations of motion in the
original frame is that the measure $\Phi$ degrees of freedom
appear only through dependence upon the scalar field $\zeta$,
Eq.(\ref{zeta}). In particular, all the equations of motion and
the solution for the connection coefficients include terms
proportional to $\partial_{\mu}\zeta$, that makes space-time non
Riemannian and all equations of motion - noncanonical. It turns
out that when working with the new metric
\begin{equation}
\tilde{g}_{\mu\nu}=e^{\alpha\phi/M_{p}}(\zeta +b_{g})g_{\mu\nu},
\label{ct}
\end{equation}
which we call the Einstein frame,
 the connection  becomes Riemannian. Since
$\tilde{g}_{\mu\nu}$ is invariant under the scale transformations
(\ref{st}), spontaneous breaking of the scale symmetry (by means
of Eq.(\ref{varphi})) is reduced in the Einstein frame to the {\it
spontaneous breakdown of the shift symmetry} $\phi\rightarrow\phi
+const.$ Notice that the Goldstone theorem generically is not
applicable in this model\cite{G}.

 Equations of motion in the Einstein frame take the standard GR
 form:
\begin{equation}
G_{\mu\nu}(\tilde{g}_{\alpha\beta})=\frac{\kappa}{2}T_{\mu\nu}^{eff}
 \label{gef}
\end{equation}
where  $G_{\mu\nu}(\tilde{g}_{\alpha\beta})$ is the Einstein
tensor in the Riemannian space-time with the metric
$\tilde{g}_{\mu\nu}$; the energy-momentum tensor
$T_{\mu\nu}^{eff}$ reads
\begin{eqnarray}
T_{\mu\nu}^{eff}=\frac{\zeta +b_{\phi}}{\zeta +b_{g}}
\left(\phi_{,\mu}\phi_{,\nu}-
\tilde{g}_{\mu\nu}X\right)+\tilde{g}_{\mu\nu}V_{eff}(\phi;\zeta,M)
-\tilde{g}_{\mu\nu}\frac{\delta\cdot b_{g}}{\zeta +b_{g}} X
\label{Tmn}
\end{eqnarray}
where
$X\equiv\frac{1}{2}\tilde{g}^{\alpha\beta}\phi_{,\alpha}\phi_{,\beta}$
\, , \, $\delta =(b_{g}-b_{\phi})/b_{g}$ \quad and
\begin{equation}
V_{eff}(\phi;\zeta ,M)=
\frac{b_{g}\left[M^{4}e^{-2\alpha\phi/M_{p}}+V_{1}\right]
-V_{2}}{(\zeta +b_{g})^{2}}. \label{Veff1}
\end{equation}

The scalar field $\zeta$  is determined by means of the constraint
which appears as the consistency condition of the equations of
motion:
\begin{eqnarray}
&&(b_{g}-\zeta)\left[M^{4}e^{-2\alpha\phi/M_{p}}+
V_{1}\right]-2V_{2}-\delta\cdot b_{g}(\zeta +b_{g})X
=0\label{constraint2}
\end{eqnarray}

Applying the constraint (\ref{constraint2}) to the $\phi$-equation
one can reduce it to the form
\begin{equation}
\frac{1}{\sqrt{-\tilde{g}}}\partial_{\mu}\left[\frac{\zeta
+b_{\phi}}{\zeta
+b_{g}}\sqrt{-\tilde{g}}\tilde{g}^{\mu\nu}\partial_{\nu}\phi\right]-\frac{2\alpha\zeta}{(\zeta
+b_{g})^{2}M_{p}}M^{4}e^{-2\alpha\phi/M_{p}} =0.
\label{phi-after-con}
\end{equation}

The effective energy-momentum tensor (\ref{Tmn}) can be
represented in a form of that of  a perfect fluid with the
following energy and pressure densities resulting from
Eqs.(\ref{Tmn}) and (\ref{Veff1}) after inserting the solution
$\zeta =\zeta(\phi,X;M)$ of Eq.(\ref{constraint2}):
\begin{equation}
\rho(\phi,X;M) =X+
\frac{(M^{4}e^{-2\alpha\phi/M_{p}}+V_{1}-\delta\cdot b_gX)^{2}
 -4\delta^{2}
b_{g}^{2}X^2}{4[b_{g}(M^{4}e^{-2\alpha\phi/M_{p}}+V_{1})-V_{2}]}
\label{rho1}
\end{equation}
\begin{equation}
p(\phi,X;M) =X- \frac{\left(M^{4}e^{-2\alpha\phi/M_{p}}+V_{1}+
\delta b_{g}X\right)^2}
{4[b_{g}(M^{4}e^{-2\alpha\phi/M_{p}}+V_{1})-V_{2}]}. \label{p1}
\end{equation}

We will proceed in a spatially flat FRW universe with  metric
$\tilde{g}_{\mu\nu}=diag(1,-a^2,-a^2,-a^2)$ filled with
homogeneous scalar field $\phi(t)$. Then the $\phi$ equation
\label{phi-after-con} reads
\begin{equation}
Q_{1}\ddot{\phi}+ 3HQ_{2}\dot{\phi}- \frac{\alpha}{M_{p}}Q_{3}
M^{4}e^{-2\alpha\phi/M_{p}}=0 \label{phi1}
\end{equation}
 where $H$ is the Hubble parameter, $\dot{\phi}\equiv d\phi/dt$
  and we have used the following notations
\begin{equation}
Q_1=(b_{g}+b_{\phi})(M^{4}e^{-2\alpha\phi/M_{p}}+V_{1})-
2V_{2}-3\delta^{2}b_{g}^{2}X \label{Q1}
\end{equation}
\begin{equation}
Q_2= (b_{g}+b_{\phi})(M^{4}e^{-2\alpha\phi/M_{p}}+V_{1})-
2V_{2}-\delta^{2}b_{g}^{2}X\label{Q2}
\end{equation}
\begin{equation}
-\frac{\alpha}{M_{p}}Q_{3}
M^{4}e^{-2\alpha\phi/M_{p}}=2[b_{g}(M^{4}e^{-2\alpha\phi/M_{p}}+V_{1})-V_{2}]\rho_{,\phi}
\label{Q3-rho-phi}
\end{equation}

It is interesting that the non-linear $X$-dependence appears here
in the framework of the underlying model without exotic terms in
the Lagrangians $L_1$ and $L_2$.  This effect follows just from
the fact that there are no reasons to choose the parameters
$b_{g}$ and $b_{\phi}$ in the action (\ref{totaction}) to be equal
in general; on the contrary, the choice $b_{g}=b_{\phi}$ would be
a fine tuning. Besides one should stress that the $\phi$
dependence in $\rho$, $p$ and in equations of motion emerges only
in the form $M^{4}e^{-2\alpha\phi/M_{p}}$ where $M$ is the
integration constant (see Eq.(\ref{varphi})), i.e. due to the
spontaneous breakdown of the scale symmetry (\ref{st}) (or the
shift symmetry $\phi\rightarrow\phi +const.$ in the Einstein
frame). Thus the above equations represent an {\it explicit
example of $k$-essence}\cite{k-essence} {\it resulting from first
principles including the scale invariance}. The system of
equations (\ref{gef}), (\ref{rho1})-(\ref{phi1})  can be obtained
from the k-essence type effective action
\begin{equation}
S_{eff}=\int\sqrt{-\tilde{g}}d^{4}x\left[-\frac{1}{\kappa}R(\tilde{g})
+p\left(\phi,X;M\right)\right] \label{k-eff},
\end{equation}
where the scalar field effective Lagrangian $p(\phi,X;M)$,
Eq.(\ref{p1}), can be represented in the form
\begin{equation}
p\left(\phi,X;M\right)=K(\phi)X+
L(\phi)X^2-\frac{[V_{1}+M^{4}e^{-2\alpha\phi/M_{p}}]^{2}}
{4[b_{g}\left(V_{1}+M^{4}e^{-2\alpha\phi/M_{p}}\right)-V_{2}]}
\label{eff-L-ala-Mukhanov}
\end{equation}
where $K(\phi)$ and $L(\phi)$ depend on $\phi$ only via
$M^{4}e^{-2\alpha\phi/M_{p}}$.  Note that  the Lagrangian
(\ref{eff-L-ala-Mukhanov}) differs from that of
Ref.\cite{k-inflation-Mukhanov} by the sign of $L(\phi)$: in our
case $L(\phi)<0$ provided the effective potential (the last term
in (\ref{eff-L-ala-Mukhanov})) is non-negative. {\it This result
cannot be removed by a choice of the parameters} of the underlying
action (\ref{totaction}) while in Ref.\cite{k-inflation-Mukhanov}
the positivity of $L(\phi)$ was an essential {\it assumption}.
This difference plays a crucial role in a number of specific
features of the scalar field dynamics, e.g. {\it the absence of
the initial singularity of the curvature is directly related to
the negative sign of} $L(\phi)$.

\section{Resolution of the Old Cosmological Constant Problem }

 The cosmological dynamics is significantly simplified if $\delta =0$.
Note that studying the cosmological constant (CC) problem we are
in fact interested in a regime where $X\to 0$. Therefore in what
it concerns the CC problem, the results of the fine tuned model
($\delta =0$ is equivalent to $b_{g}=b_{\phi}$) coincide with
those of the generic case $\delta\neq 0$. So let us consider a
spatially flat FRW cosmological model governed by
Eqs.(\ref{rho1})-(\ref{phi1}) (where we set $\delta =0$) and
$\dot{a}^{2}/a^{2}=(3M_{p})^{-2}\rho$. In the fine tuned case
under consideration,   the constraint (\ref{constraint2}) yields
\begin{equation}
\zeta =\zeta(\phi,X;M)|_{\delta =0}\equiv b_{g}-\frac{2V_{2}}
{V_{1}+M^{4}e^{-2\alpha\phi/M_{p}}},
\label{zeta-without-ferm-delta=0}
 \end{equation}
 The energy density and pressure take then the canonical form,
$\rho=\dot{\phi}^{2}/2+V_{eff}^{(0)}(\phi)$, \,
$p=\dot{\phi}^{2}/2-V_{eff}^{(0)}(\phi)$ where the effective
potential of the scalar field $\phi$ results from
Eqs.(\ref{Veff1}) and (\ref{zeta-without-ferm-delta=0})
\begin{equation}
V_{eff}^{(0)}(\phi)\equiv V_{eff}(\phi;\zeta ,M)|_{\delta =0}
=\frac{[V_{1}+M^{4}e^{-2\alpha\phi/M_{p}}]^{2}}
{4[b_{g}\left(V_{1}+M^{4}e^{-2\alpha\phi/M_{p}}\right)-V_{2}]}
\label{Veffvac-delta=0}
\end{equation}
and the $\phi$-equation (\ref{phi1}) is reduced to
$\ddot{\phi}+3H\dot{\phi}+dV^{(0)}_{eff}/d\phi =0$. Notice that
$V_{eff}^{(0)}(\phi)$ is non-negative for any $\phi$ provided
$b_{g}V_{1}\geq V_{2}$ that we will assume in what follows. We
assume also that $b_{g}>0$.

The most remarkable feature of the effective potential
(\ref{Veffvac-delta=0}) is that it is proportional to the square
of $V_{1}+ M^{4}e^{-2\alpha\phi/M_{p}}$ which is a straightforward
consequence of our basic assumption that $L_1$ and $L_2$ are
independent of the measure fields. Due to this, as $V_{1}<0$ and
$V_{2}<0$, {\it the effective potential has a minimum where it
equals zero automatically}, without any further tuning of the
parameters $V_{1}$ and $V_{2}$. This occurs in the process of
evolution of the field $\phi$ at the value of $\phi =\phi_{0}$
where $V_{1}+ M^{4}e^{-2\alpha\phi_{0}/M_{p}}=0$. This means that
the universe evolves into the state with zero cosmological
constant without any additional tuning of the parameters  and
initial conditions.

 If such type of the structure for the scalar field potential in a
conventional (non TMT) model would be chosen "by hand" it would be
a sort of fine tuning. But in our TMT model it is not the starting
point, {\it it is rather a result} obtained in the Einstein frame
of TMT models with spontaneously broken global scale symmetry
including the shift symmetry $\phi\rightarrow \phi +const$. Note
that the assumption of scale invariance is not necessary for the
effect of appearance of  the perfect square in the effective
potential in the Einstein frame and therefore for the described
mechanism of disappearance of the cosmological constant, see
Ref.\cite{GK1}. To provide the global scale invariance (\ref{st}),
the prepotentials $V_1$ and $V_2$ enter in the action
(\ref{totaction}) with factor $e^{2\alpha\phi/M_p}$. However, if
quantum effects (considered in the original frame) break the scale
invariance of the action via modification of existing
prepotentials or by means of generation of other prepotentials
with arbitrary $\phi$ dependence (and in particular a "normal"
cosmological constant term $\int \tilde{\Lambda}\sqrt{-g}d^4x$),
this cannot change the result of TMT that the effective  potential
generated in the Einstein frame is proportional to a perfect
square.

At the first glance this effect contradicts the Weinberg's no-go
theorem\cite{Weinberg1} which states that there cannot exist a
field theory model where the cosmological constant is zero without
fine tuning. However one can show that conditions of the
Weinberg's theorem do not hold in our model. Note that $\zeta
=\zeta_{0}(\phi)$, Eq.(\ref{zeta-without-ferm-delta=0}), becomes
singular
\begin{equation}
|\zeta|\approx\frac{2|V_2|}{|V_{1}+
M^{4}e^{-2\alpha\phi/M_{p}}|}\rightarrow\infty \qquad as \qquad
\phi\rightarrow\phi_{0}. \label{zeta-singular}
\end{equation}
  In this limit the effective potential
 (\ref{Veff1}) (or
 Eq.(\ref{Veffvac-delta=0})) behaves as
$V_{eff}\approx\frac{|V_2|}{\zeta^2}$. Thus, disappearance of the
cosmological constant occurs in the regime where
$|\zeta|\equiv|\Phi|/\sqrt{-g}\rightarrow\infty$. In this limit,
the dynamical role of the terms of the Lagrangian $L_2$ (coupled
with the measure $\sqrt{-g}$) in the action (\ref{totaction})
becomes negligible in comparison with the terms of the Lagrangian
$L_1$ (see also the general form of the action (\ref{S})).  It is
evident that {\bf the limit of the TMT action (\ref{S}) as
$|\zeta|\rightarrow\infty$ is opposite to the conventional field
theory (with only measure $\sqrt{-g}$) limit of the TMT action}.
From the point of view of TMT, this is {\bf the answer to the
question why the old CC problem cannot be solved (without fine
tuning) in theories with only the measure of integration}
$\sqrt{-g}$ {\bf in the action}.

Recall that one of the basic assumptions of the Weinberg's no-go
theorem is that all fields in the vacuum must be constant. This is
also assumed for the metric tensor, components of which  in the
vacuum must be {\bf nonzero} constants. However, this is not the
case in the underlying TMT action (\ref{totaction}) defined in the
original (non Einstein) frame if we ask what is the metric tensor
$g_{\mu\nu}$ in the $\Lambda =0$ vacuum . To see this let us note
that in the Einstein frame all the terms in the cosmological
equations are regular. This means that the metric tensor in the
Einstein frame $\tilde{g}_{\mu\nu}$ is always well defined,
including the $\Lambda =0$ vacuum state $\phi =\phi_{0}$ where
$\zeta$ is infinite. Taking this into account and using the
transformation to the Einstein frame (\ref{ct}) we see that {\bf
all components of the metric in the original frame $g_{\mu\nu}$ go
to zero overall in space-time as $\phi$ approaches the $\Lambda
=0$ vacuum state}:
\begin{equation}
g_{\mu\nu}\sim \frac{1}{\zeta}\sim V_{1}+
M^{4}e^{-2\alpha\phi/M_{p}}\rightarrow 0 \quad (\mu,\nu =0,1,2,3)
\quad as \quad \phi\rightarrow\phi_{0}. \label{g-degen}
\end{equation}
This result shows that the Weinberg's analysis based on the study
of the trace of the energy-momentum tensor misses any sense in the
case $g_{\mu\nu}=0$.

It follows immediately from (\ref{g-degen}) that $\sqrt{-g}$ tends
to zero like $\sqrt{-g}\sim \zeta^{-2}$ as
$\phi\rightarrow\phi_{0}$. Then the definition (5) implies that
the integration measure $\Phi$ also tends to zero but rather like
$\Phi\sim \zeta^{-1}$ as $\phi\rightarrow\phi_{0}$. Thus both the
measure $\Phi$ and the measure $\sqrt{-g}$ become degenerate  in
the $\Lambda =0$ vacuum state $\phi =\phi_{0}$. However
$\sqrt{-g}$ tends to zero more rapidly than $\Phi$.

As we have discussed in detail\cite{GK1}, with the original set of
variables used in the underlying TMT action (\ref{S}),
(\ref{totaction}) it is very hard or may be even impossible to
display the physical meaning of TMT models. One of the reasons is
that in the framework of the postulated need to use the Palatini
formalism, the original metric $g_{\mu\nu}$ and connection
$\Gamma^{\alpha}_{\mu\nu}$ appearing in the underlying TMT action
describe a non-Riemannian space-time. The transformation to the
Einstein frame (\ref{ct}) enables to see the physical meaning of
TMT because the space-time becomes Riemannian in the Einstein
frame. Now we see that the transformation to the Einstein frame
(\ref{ct}) plays also the role of a {\bf regularization of the
space-time metric}: the singular behavior of the transformation
(\ref{ct}) as $\phi\approx\phi_{0}$ compensates the disappearance
of the original metric $g_{\mu\nu}$ in the vacuum $\phi
=\phi_{0}$. As a result of this the metric in the Einstein frame
$\tilde{g}_{\mu\nu}$ turns out to be well defined in all physical
states including the $\Lambda =0$ vacuum state.

\section{
Absence of Initial Singularity of the Curvature and Inflationary
Cosmology
 with Graceful Exit to $\Lambda =0$ Vacuum}

Let us start from the analysis of Eq.(\ref{phi1}). The interesting
feature of this equation is that
 each of the factors $Q_{i}(\phi,X)$ \, ($i=1,2,3$) can get
 to zero and this effect depends on the range of the parameter
 space chosen.

\begin{figure}[htb]
\begin{center}
\includegraphics[width=12.5cm,height=9cm]{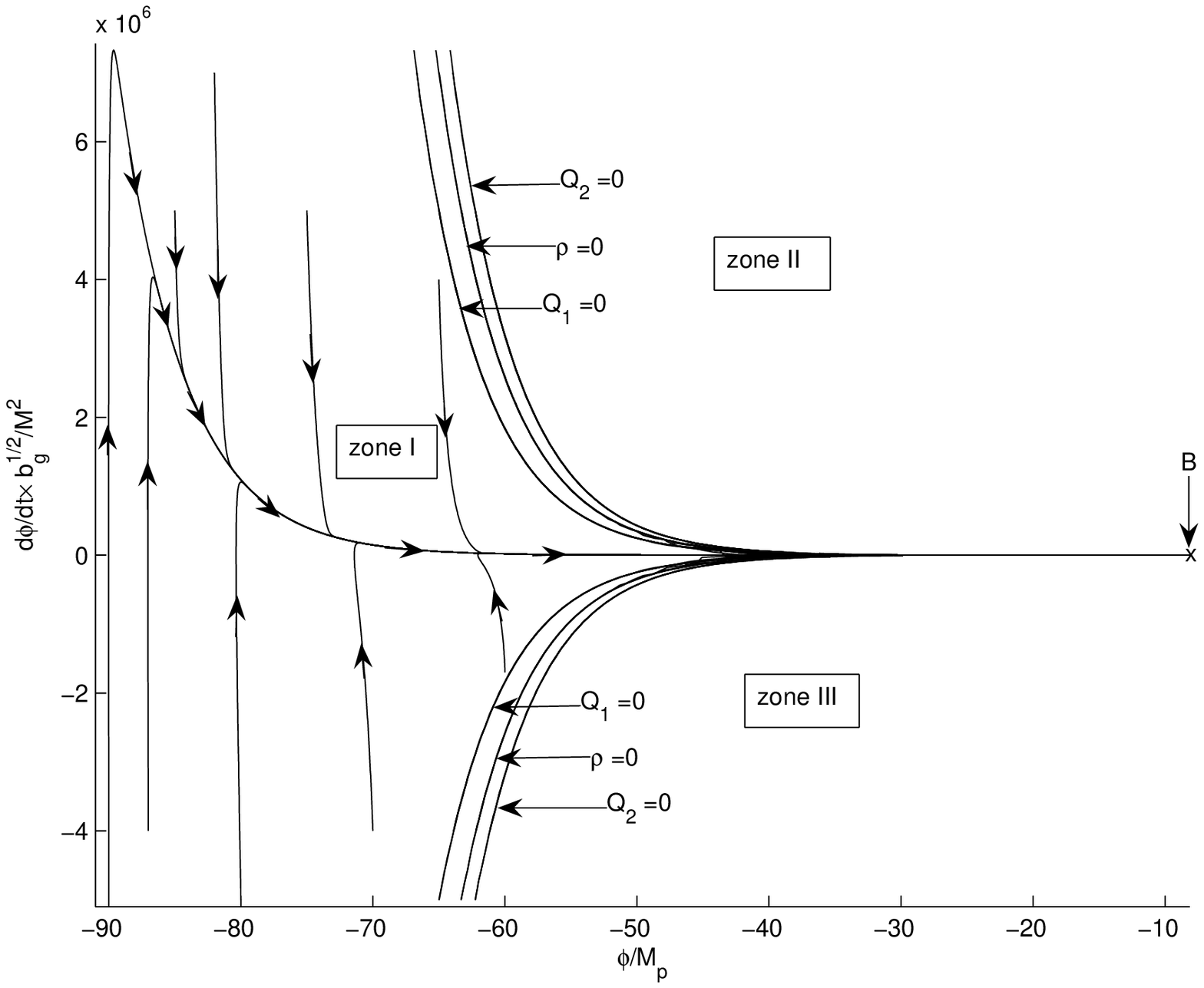}
\end{center}
\label{fig9}
\end{figure}

For $Q_{1}\neq 0$, Eq. (\ref{phi1}) results in the well known
equation\cite{Vikman}
\begin{equation}
\ddot{\phi}+
\frac{\sqrt{3\rho}}{M_p}c_s^2\dot{\phi}+\frac{\rho_{,\phi}}{\rho_{,X}}
=0, \label{phantom-phi}
\end{equation}
where $c_s$ is the effective sound speed of perturbations
\begin{equation}
c_s^2=\frac{p_{,X}}{\rho_{,X}}=\frac{Q_2}{Q_1} \qquad and \qquad
\frac{\rho_{,\phi}}{\rho_{,X}}=- \frac{\alpha}{M_{p}}\cdot
\frac{Q_3}{Q_1}\cdot M^{4}e^{-2\alpha\phi/M_{p}}, \label{Q-rho-p}
\end{equation}
$\rho$ and $p$ are defined by Eqs.(\ref{rho1}), (\ref{p1}) and
$Q_i$ $(i=1,2,3)$ - by Eqs.(\ref{Q1})-(\ref{Q3-rho-phi}).

It follows from the definitions of $Q_1$ and $Q_2$ that $c_s^2>1$
when $Q_1>0$ and $X>0$ (that implies $Q_2>0$).  However $c_s^2<1$
when $Q_1<0$ and $Q_2<0$.

In the model with $V_1<0$ and $V_2<0$,  the structure of the phase
plane is presented in the figure above for the following set of
the parameters: $\alpha =0.2$, $\delta =0.1$, $V_1=-30$,
$V_2=-50$. With such a choice of the parameters the following
condition is satisfied
\begin{equation}
(b_g+b_{\phi})V_1-2V_2>0 \label{cond-for-phase-structure}
\end{equation}
that determines the structure of the phase plane.
 By the two branches
of the line $Q_1=0$, the phase plane is divided into three large
dynamically disconnected zones. In the most part of two of them
(II and III), the energy density is negative ($\rho <0$). In the
couple of regions between the lines $Q_1=0$ and $\rho =0$, $\rho
>0$ but $Q_1<0$.

The zone I of the phase plane (where $Q_1>0$, $\rho
>0$ and $Q_2>0$) is of a great cosmological interest:
All the phase
 curves start with very steep approach to an attractor and they arrive
 the attractor  asymptotically.  One
 can show that the attractor does not intersect the line
 $Q_1=0$. In the neighborhood of the line $Q_1=0$ all the phase curves
exhibit  a {\it repulsive} behavior from this line. In other
words, the shape of two branches of $Q_1=0$ do not allow a
classical dynamical continuation of the phase curves to the past
in time without crossing the classical barrier formed by the line
$Q_1=0$. This is true for all finite values of the initial
conditions $\phi_{in}$, $\dot{\phi}_{in}$ in zone I. The
cosmological evolution corresponding to the phase curves with
$\phi_{in}\ll -M_p$ is a power law inflation which ends with the
graceful exit to a zero CC vacuum state without fine tuning (point
B in the figure marks the corresponding oscillatory regime).

In two regions between the lines $Q_1=0$ and $Q_2=0$ that we refer
for short as ($Q_1\rightarrow Q_2$)-regions,, the squared sound
speed of perturbations is negative, $c_s^2<0$. This means that on
the right hand side of the classical barrier $Q_1=0$, the model is
absolutely unstable. Moreover, this pure imaginary sound speed
becomes infinite in the limit $Q_1\rightarrow 0-$. Thus the
branches of the line $Q_1=0$ divide zone I (of the classical
dynamics) from the ($Q_1\rightarrow Q_2$)-regions where the
physical significance of the model is unclear. Note that the line
$\rho =0$ divides the ($Q_1\rightarrow Q_2$)-regions into two
subregions with opposite signs of the classical energy density.

Thus the structure of the phase plane yields a conclusion that
 the starting point of the classical history in the phase
plane can be only in zone I and {\bf the line $Q_1 =0$ is the
limiting set of points where the classical history might begin}.
For any finite initial values of $\phi_{in}$ and $\dot{\phi}_{in}$
at the initial cosmic time $t_{in}$, the  duration $t_{in}-t_{s}$
of the continuation of the evolution into the past up to the
moment $t_{s}$ when the phase trajectory arrives the line $Q_1
=0$, is finite. Note that $\dot{\phi_s}=\dot{\phi}(t_s)$, the
energy density $\rho_{s}=\rho(t_{s})$ and the pressure
$p_{s}=p(t_{s})$ are finite in all points of the line $Q_1 =0$
with finite  $\phi_s=\phi(t_s)$. One can also show that the strong
energy condition is satisfied in regions of zone I close to the
line $Q_1 =0$ including the line itself. The detailed analysis
yields the conclusion that the singularity as $t\to t_{s}$ is a
sort of sudden singularities studied by Barrow\cite{Barrow1} on
purely kinematic grounds: $\dot{\phi}\propto \sqrt{t-t_s}$, the
scalar curvature is finite but its time derivative $\dot{R}\propto
-a^{-1}d^3a/dt^3\propto -(t-t_s)^{-1/2}\to -\infty$.

\section{Crossing the Phantom Divide }

 Equations
$Q_{i}(\phi,\dot\phi)=0$ \, ($i=1,2,3$) \, determine lines in the
phase plane $(\phi,\dot\phi)$. In terms of a mechanical
interpretation of Eq.(\ref{phi1}), the change of the sign of
$Q_{1}$ can be treated as the change of the mass of "the
particle". Therefore one can think of situation where "the
particle" climbs up in the potential with acceleration. It turns
out that when the scalar field is behaving in this way, the flat
FRW universe may undergo a super-acceleration.

\begin{figure}[htb]
\begin{center}
\includegraphics[width=12.5cm,height=8.0cm]{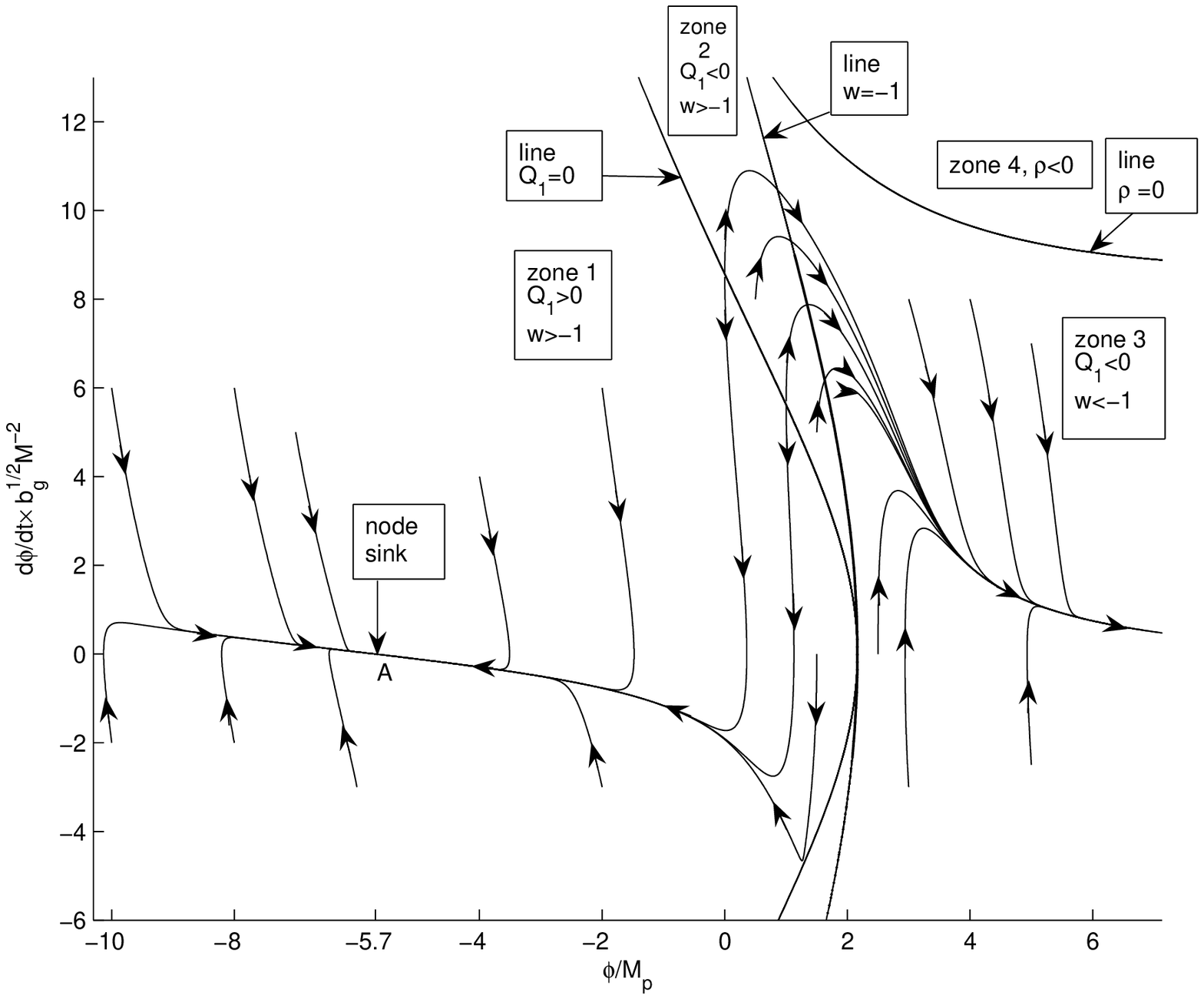}
\end{center}
\label{fig17}
\end{figure}

With simple algebra one can see that the following "sign rule"
holds for the equation-of-state: $sign(w+1)=sign (Q_2)$. Therefore
if in the one side from the line $Q_2(\phi,\dot\phi)=0$ \, $w>-1$
then in the other side $w<-1$. To incorporate the region of the
phase plane where $w<-1$ into the cosmological dynamics one should
provide that the line $Q_2(\phi,\dot\phi)=0$ lies in a dynamically
permitted zone. As we have seen in Sec.4, the condition
(\ref{cond-for-phase-structure}) disposes the line
$Q_2(\phi,\dot\phi)=0$ in the physically unacceptable region where
$\rho<0$. It turns out that if the opposite condition holds
\begin{equation}
(b_g+b_{\phi})V_1-2V_2<0 \label{cond-for-phase-structure-1}
\end{equation}
then the line $Q_2(\phi,\dot\phi)=0$ may be in a dynamically
permitted zone.

There are a lot of  sets of parameters providing the $w<-1$ phase
in the late universe. For example we are demonstrating here this
effect with the following set of the parameters of the underlying
action(\ref{totaction}): $\alpha =0.2$, $V_{1}=10M^{4}$ and
$V_{2}=9.9b_{g}M^{4}$, $\delta =0.1$. The results of the numerical
solution are presented in the figure above. The phase plane is
divided into two regions by the line $\rho =0$. The region $\rho
>0$ is divided into two dynamically disconnected regions by the
line $Q_{1}(\phi,\dot\phi)=0$. To the left of the line
$Q_{1}(\phi,\dot\phi)=0$ - zone 1 where $Q_{1}>0$. In the zone 1,
there is a node sink $A(-5.7M_{p},0)$. Among trajectories
converging to node $A$ there are also trajectories corresponding
to a power low inflation of the early universe. Note that the same
mechanism of the sudden initial singularity we have discussed in
Sec.4, holds also here for solutions whose phase curves are
located in zone 1.

In the region to the right of the line $Q_{1}(\phi,\dot\phi)=0$,
all phase curves approach the attractor which in its turn
asymptotically (as $\phi\rightarrow\infty$) takes the form of the
straight line $\dot\phi =0$. This region is divided into two zones
by the line $Q_2(\phi,\dot\phi)=0$. In all points of this line
$w=-1$. In zone 2, i.e. between the lines $Q_1=0$ and $w=-1$, the
equation-of-state $w>-1$ and the sound speed $c_s^2<0$. Therefore
in zone 2 the model is absolutely unstable and has no any physical
meaning. In zone 3, i.e. between the line $w=-1$ and the line
$\rho =0$, the equation-of-state $w<-1$ and the sound speed
$c_s^2>0$.  The main features of the solution of the equations of
motion are the following (part of them, for a particular choice of
the initial data $\phi_{in}=M_{p}$, $\dot\phi_{in}
=9M^4/\sqrt{b_g}$, one can see in the figure below): 1) $\phi$
slowly increases in time; 2) the energy density $\rho$ slowly
increases approaching a constant;
 3) in zone 3, $w$ becomes
less than $-1$ and after achieving a minimum ( for the initial
data chosen, this minimum is $w\approx -1.2$) it then increases
asymptotically approaching $-1$ from below. Using the
classification of Ref.\cite{Vikman} of conditions for
 the dark energy to evolve from the state with $w>-1$ to the
phantom state, we see that transition of the phase curves from
zone 2 (where $w>-1$) to the phantom zone 3 (where $w<-1$) occurs
under the conditions  $p_{,X}=0$, $\rho_{,X}\neq 0$, $X\neq 0$.
Qualitatively the same behavior one observes for all initial
conditions $(\phi_{in},\dot{\phi}_{in})$ given in the zone 2.

\begin{figure}[htb]
\begin{center}
\includegraphics[width=12.5cm,height=6.0cm]{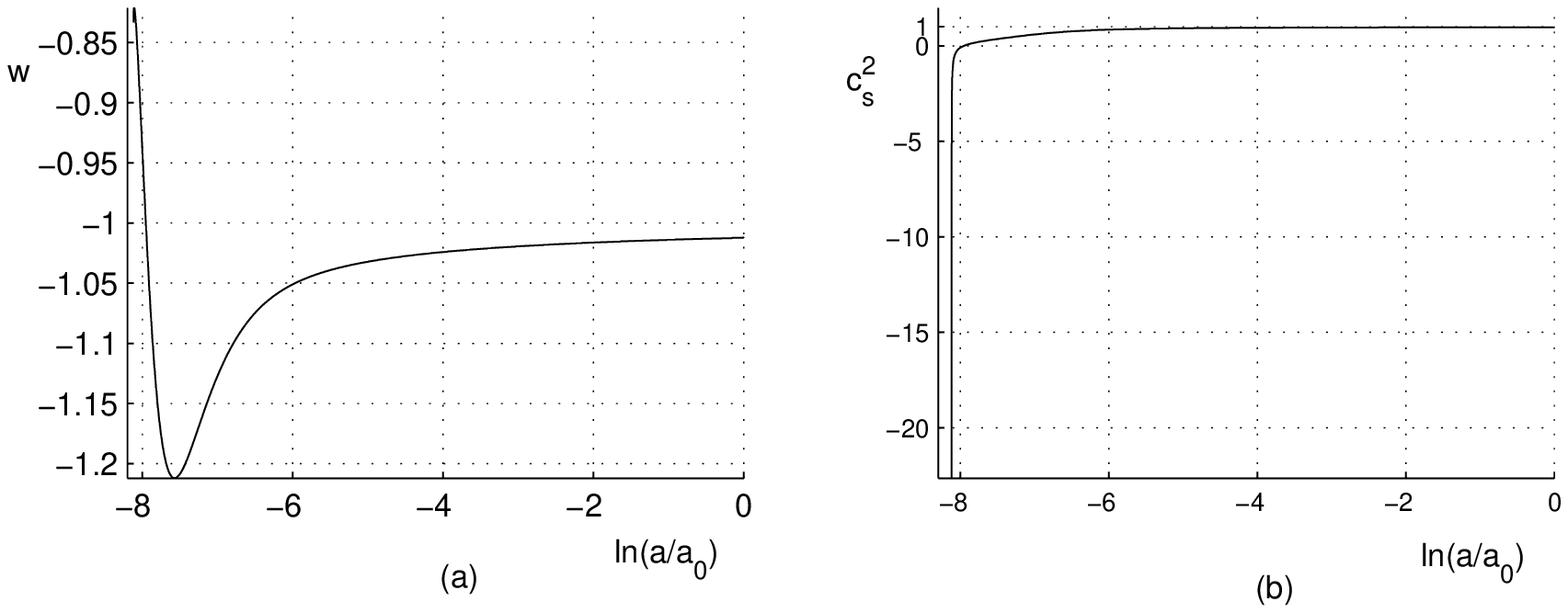}
\end{center}
 \label{fig20}
\end{figure}

%\clearpage % delate this line


\begin{thebibliography}{B-B} %%%%%
\medskip
\begin{footnotesize} % do not change this line !!!

\bibitem{GK1} E.I. Guendelman and A.B. Kaganovich, {\it Phys.
Rev.} {\bf D57}, 7200 (1998); ibid. {\bf D60}, 065004 (1999); {\it
Int. J. Mod. Phys.} {\bf A17}, 417 (2002); {\it Mod. Phys. Lett.}
{\bf A17}, 1227 (2002); {\it Int.J.Mod.Phys.} {\bf A21}, 4373
(2006);
 E.I. Guendelman, {\it Mod.
Phys. Lett.} {\bf A14}, 1043 (1999).

\bibitem{GK2-Higgs} E.I. Guendelman and A.B. Kaganovich,
hep-th/0603150.

\bibitem{GK3} E.I. Guendelman and A.B. Kaganovich, gr-qc/0607111,
to appear in Phys. Rev. D.


\bibitem{G} E.I. Guendelman, {\it Class. Quant. Grav.} {\bf 17}, 361
(2000).


\bibitem{k-essence} C. Armendariz-Picon, V. Mukhanov and P.J. Steinhardt,
{\it Phys.Rev.Lett.} {\bf 85} 4438, 2000.



\bibitem{k-inflation-Mukhanov}
C. Armendariz-Picon, T. Damour and V. F. Mukhanov, {\it
Phys.Lett.} B{\bf 458}, 209 (1999).

\bibitem{Weinberg1}
S. Weinberg, {\it Rev. Mod. Phys.} {\bf 61}, 1 (1989).

\bibitem{Barrow1}
J.D. Barrow, {\it Class.Quant.Grav.} {\bf 21}, L79 (2004); ibid.
{\bf 21}, 5619 (2004).

\bibitem{Vikman}
A. Vikman, {\it Phys.Rev.} {\bf D71}, 023515 (2005).


\end{footnotesize} % do not change this line !!
\end{thebibliography}
\end{document}